\documentclass[structabstract]{aa}  
\usepackage{graphicx}
\usepackage{epstopdf}
\usepackage{color}
\usepackage{longtable}

\usepackage{txfonts}
\usepackage{natbib}

\def\gr{$\gamma$-ray}
\def\mrk{Mrk~501}

\begin{document}

\title{Very hard gamma-ray emission from a flare of Mrk 501}

    \author{A.~Neronov
          \inst{1},
          D.~Semikoz\inst{2,3}
          \and
          A.M.~Taylor
          \inst{1}
          }

   \institute{ISDC Data Centre for Astrophysics, Ch. d'Ecogia 16, 1290, Versoix, Switzerland \\
              \email{Andrii.Neronov@unige.ch}
         \and
             APC, 10 rue Alice Domon et Leonie Duquet, F-75205 Paris Cedex 13, France \and
Institute for Nuclear Research RAS, 60th October Anniversary prosp. 7a, Moscow, 117312, Russia
\\
             \email{Dmitri Semikoz <dmitri.semikoz@apc.univ-paris7.fr>}
             }

\abstract
{}
{
We investigate the peculiar properties of a large TeV \gr\ flare from Mrk~501 detected during the 2009 multiwavelength campaign. 
}
{
We identify the counterpart of the flare in the \textit{Fermi} LAT telescope data and study its spectral and timing characteristics. 
}
{
A strong order-of-magnitude increase of the very-high-energy \gr\ flux during the flare was not accompanied by an increase in the X-ray flux, so that the flare was one of the ``orphan''-type TeV flares observed in BL Lacs. The flare lasted about 1 month at energies above 10~GeV. The flaring source spectrum in the 10-200~GeV range was very hard, with a photon index $\Gamma=1.1\pm 0.2$, harder than that observed in any other blazar in the \gr\ band. No simultaneous flaring activity was detected below 10~GeV.  Different variability properties of the emission below and above 10~GeV indicate the existence of two separate components in the spectrum. We investigate possible explanations of the very hard flaring component.  We consider, among others, the possibility that the flare is produced by an electromagnetic cascade initiated by very-high-energy \gr s in the intergalactic medium. Within such an interpretation, peculiar spectral and temporal characteristics of the flare could be explained if the magnetic field in the intergalactic medium is of the order of $10^{-17} - 10^{-16}$~G. 
}
{}

\maketitle

\section{Introduction}
Variable \gr\ emission from BL Lacs is most commonly interpreted as Doppler-boosted inverse Compton (IC) emission by high-energy electrons in relativistic jets aligned with the line of sight. In the Synchrotron-self-Compton (SSC)  model, the seed photons for the IC scattering are provided by the lower energy (radio-to-X-ray) synchrotron emission produced by the same electrons. Since both the synchrotron and IC emissions are produced by the same electrons, the variability of the \gr\ emission is expected to be accompanied by a similar variability in the X-ray emission. However, this is not always the case. Strong ``orphan''  \gr\ flares in the TeV energy band are sometimes observed without accompanying X-ray synchrotron flares \citep{krawczynski04,daniel05}. The nature of these orphan flares is not clear. Different mechanisms have been proposed to explain these orphan flares in the context of leptonic \citep{kusunose06} or hadronic \citep{bottcher05} models of blazar emission.

In the framework of the SSC model for the \gr\ emission, the emission spectrum is expected to be characterized by a photon index $\Gamma\ge 1.5$ ($\Gamma$ is defined as the slope of the differential spectrum $dN_\gamma/dE\sim E^{-\Gamma}$) in the energy range where electron cooling via synchrotron and/or IC energy losses is efficient. The limiting value of $\Gamma=1.5$ is considered as a lower bound on $\Gamma$ in the analysis of VHE \gr\ flux attenuation through interactions with Extragalactic Background Light (EBL) \citep{1ES_nature,0229_HESS,1ESSS_HESS}. At the same time, harder than $\Gamma=1.5$ spectra could, in principle, be produced in the VHE \gr\ band as a result of absorption of \gr s in $e^+e^-$ pair production inside the source \citep{aharonian08}, due to the presence of low-energy cut-offs or hardenings in the electron distribution \citep{katarzinski06,bottcher08} or the formation of narrow energy distribution of emitting particles in the source \citep{aharonian02}. The observation of blazars with intrinsically hard \gr\ spectra, harder than $\Gamma=1.5$, would have important consequences for EBL studies and for the physical models describing the \gr\ emission.

In this letter we consider the spectral and timing characteristics of a large \gr\ flare of \mrk\ observed during a multi-wavelength campaign in 2009 \citep{fermi_paper}. The flare was remarkable both because it was an ``orphan'' flare (the factor of 10 increase in the VHE \gr\ flux was not accompanied by a similar simultaneous increase in the X-ray flux) and because the spectrum of the \gr\ emission during the flare was much harder than the spectrum of the quiescent emission from the source.  We show that the spectral and timing characteristics of the source during the flare suggest the existence of two distinct components in the \gr\ emission: a non-flaring one, characterized by a softer photon index $\Gamma\simeq 1.8$, close to the time-averaged photon index of the source,  and a flaring one, with an extremely hard spectrum $\Gamma\simeq 1.1$. We discuss the possible physical origins of this hard flaring component.
\vskip0.5cm

\section{Fermi data selection and data analysis}

For our analysis we used the publicly available data collected by the Large Area Telescope (LAT) on board the {\it Fermi} satellite \citep{atwood09} between August 4, 2008 and March 15, 2011. We processed the ``PASS 6'' data using the {\it Fermi} Science Tools\footnote{\tt http://fermi.gsfc.nasa.gov/ssc/data/analysis/scitools/}. We filtered the entire data set with {\it gtselect} and {\it gtmktime} tools following the recommendations of the {\it Fermi} team\footnote{\tt http://fermi.gsfc.nasa.gov/ssc/data/analysis/} and  retained only events belonging to the class 4, which are most confidently identified with \gr s, as described by \citet{fermi_background}.   To estimate the flux from the photon counts in 30-400 GeV band we used the {\it gtexposure} tool.  The spectral analysis was done using the unbinned likelihood method \citep{likelihood}, taking into account all the sources from the 1-st year {\it Fermi} catalog \citep{fermi_catalog} within $15^\circ$ from \mrk\ as well as an additional bright source at the position of a quasar B1638+3952, not listed in the catalog.
\vskip0.5cm

\begin{figure}
\includegraphics[width=\linewidth]{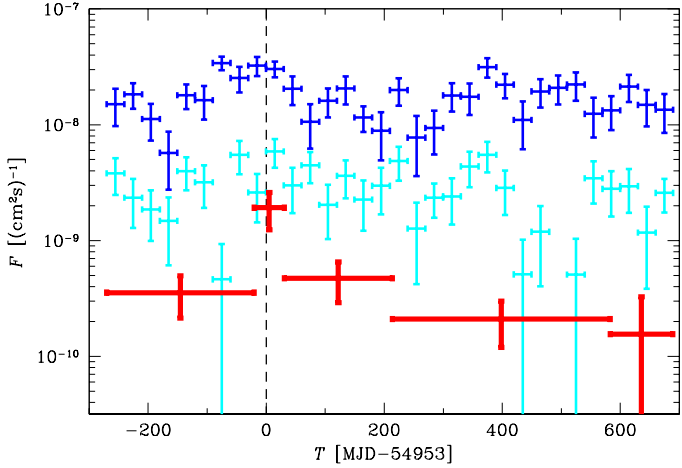}
\caption{Lightcurves of \mrk\ in the 30-300~GeV (thick red data points, errorbars are 68\% confidence levels), 3-30~GeV (thin cyan data points) and 0.3-3~GeV (dark blue thin data points) bands. The vertical dashed line marks the start of the flare detected in the TeV band by Veritas \citep{fermi_paper}.}
\label{fig:lc}
\end{figure}

{\it Timing analysis.}
The long-term lightcurves of \mrk\ for the whole period of observations are shown in Fig. \ref{fig:lc} for three different energy bands. For the highest energy band 30-400~GeV, the lightcurve was produced using the aperture photometry method, using time bins with an equal signal-to-noise ratio $S/N=3$. To estimate the diffuse background level in this energy bin we used a nearby source-free region of radius $3^\circ$ centered on Ra$=256.21^\circ$, Dec$=34.97^\circ$. In the lower energy bands the binning was homogeneous in time, with time bins of 30 days, produced using the likelihood analysis.

A flare with a flux increase by a factor of $\sim 10$ was readily identified in the lightcurve in the 30-400~GeV energy band. The flare period includes the Veritas flare reported by \citet{fermi_paper}. The moment of onset of the Veritas flare is marked in Fig. \ref{fig:lc} by the vertical dashed line.  The flare could not be identified in the lower energy lightcurves shown in Fig. \ref{fig:lc}. In principle, if the flux enhancement during the flare was moderate and the flare duration was much shorter than 1 month,  the flare signal would have been ``diluted'' in the lightcurves binned using the 30 day binning. We have verified that reducing the size of the time bins down to several days (the time scale on which the source could be significantly detected at lower energies) does not reveal the flare in the lower energy lightcurves. 

\begin{figure}
\includegraphics[width=\linewidth]{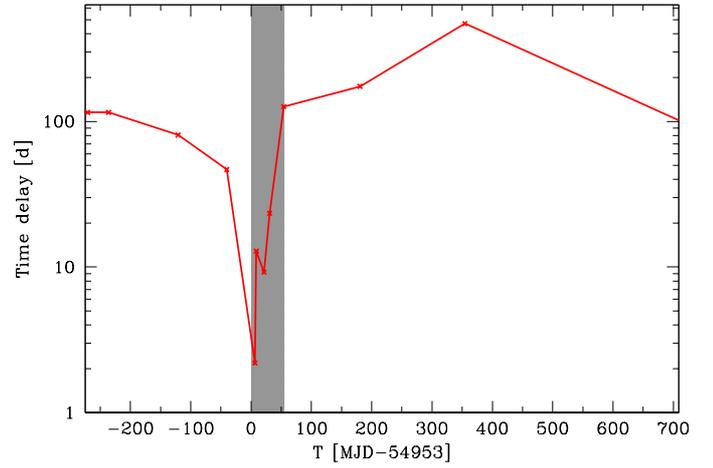}
\caption{Waiting time between subsequent photons in 100-400 GeV band. The grey shaded interval marks the time of the flare.}
\label{fig:timedelay}
\end{figure}

The duration of the flare in the VHE band was not constrained by the Veritas observations \citep{fermi_paper}, since their observations stopped 3 days after the onset of the flare, when the source was still in the flaring state. In spite of the low statistics of the VHE signal in LAT, the LAT data could be used to estimate the duration of the flare. Fig.  \ref{fig:timedelay} shows the time delay between subsequent photons in the 100-400 GeV band. During the flare the time delay between subsequent photons dropped below 10 days for about one month. During the rest of the {\it Fermi} exposure the $E>100$~GeV photon count rate from the source was about 1 photon per 100~days. To estimate the duration of the flare we divide all the 100-400~GeV photons into ``flare signal'' and ``background''. The ``background'' count rate is $\le 10^{-2}$~ph/day. The time interval $\sim 55$~d following MJD 54953 has 5 photons, four of which we ascribe to the ``flare signal'' (the probability for a background photon to come in the $\sim 50$~d interval is 0.5). The $\sim 55$~d time scale can be considered as an estimate of the flare duration. The uncertainty of this estimate is by a factor of $\sim 2$. Indeed, if the flare duration was $\sim 100$~d, instead of $\sim 50$~d,  the probability for all four ``flare signal'' photons to come in the first half of the time interval would be $\sim 9\%$, so that the $100$~d flare duration is excluded at $\simeq 90\%$ confidence level. The waiting time between subsequent photons drops down to $\sim 2$~d at the peak of the flare. Just before the peak no photons were detected during the previous $40$~d. This constrains the rise time of the flare. Assuming that the flux level comparable to the peak flux was sustained over a rise time interval $T_r$, one could estimate an upper limit on $T_r$.  The probability that no photons are detected during a time interval $\sim 8$~d before the peak is $\sim 2^{-4}\simeq 6$\%, so that $T_r>10$~d is ruled out at $>90$\% confidence level.

\vskip0.5cm

{\it Spectral analysis.}
The timing behaviour of the source below and above 10~GeV is different: the flare appears most prominently at the highest energies ($>30$~GeV) and is not detectable at low energies.  Taking this into account we have analyzed the low- and high-energy part of the LAT energy band separately. 
We have extracted the spectra in the 0.3-10~GeV and 10-400~GeV bands for the time period of the flare, between $T_0=\mbox{MJD\ } 54953$ and $T_0+30$~d, corresponding to the highest count rate of $E>100$~GeV photons, see Fig. \ref{fig:timedelay}. We assume that in each energy band the spectrum is well described by a simple powerlaw $dN_\gamma/dE=A\left(E/E_0\right)^{-\Gamma}$. 

\begin{figure}
\includegraphics[width=\linewidth]{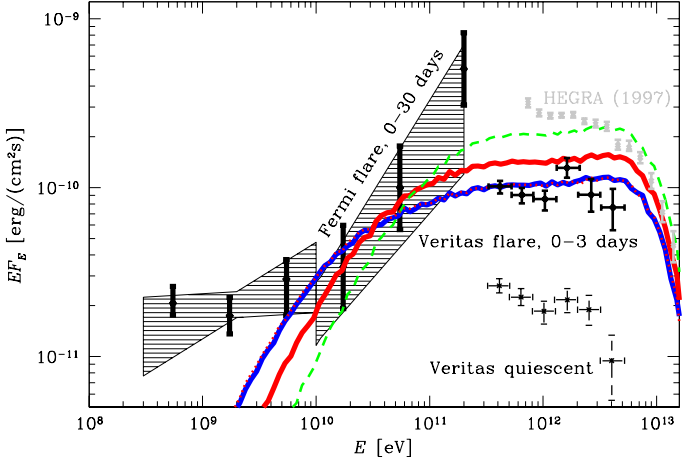}
\caption{{\it Fermi} spectrum of \mrk\ for the 30 day period during the flare (shared area,  68\% confidence level of normalization and slope of the spectrum).  Data points in the TeV band show Veritas measurements \citep{fermi_paper} of the spectrum during the first 3 days of the flare (solid lines) and in the quiescent state (dashed lines).  Model curves show emission from electromagnetic cascade initiated by 100~TeV \gr s in the intergalactic medium. The solid thick red curve is for intergalactic magnetic field $3\times 10^{-17}$~G with a correlation length of 1~Mpc. Dashed curves are for $B=10^{-16}$~G  (upper curve) and $B=10^{-17}$~ G (lower curve). Grey data points show the spectrum of the historically brightest source state in 1997 measured by CAT telescope \citep{hegra}.}
\label{fig:spectrum}
\end{figure}

The results of the likelihood analysis of the spectrum in the two energy bands are shown in Fig. \ref{fig:spectrum}. One can see that the slopes of the spectrum below and above 10~GeV are significantly different. 
Below 10~GeV the likelihood analysis gives the slope $\Gamma_{\rm le}=1.8\pm 0.2$, while  above 10~GeV the spectrum is much harder, with $\Gamma_{\rm he}=1.1\pm 0.2$. We note that the highest energy photon from the source detected by LAT during the flare period had an energy $E_{max}=188$~GeV. Taking this into account we show the powerlaw model for the source spectrum only up to 200~GeV in Fig. \ref{fig:spectrum}.

\begin{figure}
\includegraphics[width=0.8\linewidth]{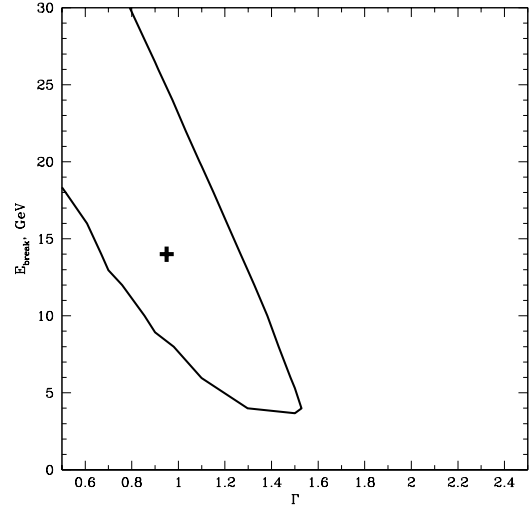}
\caption{Best fit values  (cross) and the 68\% confidence contour for the $E_{break}$, $\Gamma$ parameters of the broken powerlaw model of the flare spectrum.}
\label{fig:Ebr-k}
\end{figure}

The fact that most of the highest energy photons from the  source come in a narrow time window around the time of the flare  was already noticed by \citet{fermi_paper}. This peculiarity of the highest energy photon timing was found to lead to the hardest source spectrum during the 30~d flare period, with an average slope of $\Gamma=1.64\pm 0.09$ in the 0.2-400~GeV band. Our analysis of the spectral variability details during the flare suggests that the flaring activity appears only at the highest energies. Taking this into account, we fit the spectrum extracted from narrow energy bins (black thick data points in Fig. \ref{fig:spectrum}) with a broken powerlaw, instead of the simple powerlaw model. We find that the use of the broken powerlaw model improves the quality of the fit, with an F-test probability of the chance fit improvement at the level of 8\%. The best fit value of the break energy is found fitting the data by the broken powerlaw model is $E_{\rm break}=14$~GeV. The quality of the data is not sufficient for limiting the break energy from above, only a lower bound $E_{\rm break}>3$~GeV could be found (Fig. \ref{fig:Ebr-k}). 
If the break energy is left free during the fitting, the range of allowed values of $\Gamma_{\rm he}$ is wider, so that the lower bound on $\Gamma_{\rm he}\le 1.5$.

The difference in the spectral slopes and in the timing behaviour at low and high energies point to the presence of two separate components in the emission from the source. The ``soft'' component which appears below $\simeq 10$~GeV is steady while the ``hard'' component, visible mostly above 30~GeV, is variable during the flare. Two additional arguments which support the hypothesis for the existence of a slow-variable soft component are the closeness of the slope of the spectrum in the 0.3-10~GeV band to the time-average slope  $\Gamma\simeq 1.85$ \citep{fermi_catalog} and the consistency of a powerlaw extrapolation of the low-energy spectrum to the VHE band with the quiescent spectrum of the source measured by Veritas and MAGIC (shown by the dashed data points in Fig. \ref{fig:spectrum}). 

\section{Discussion: origin of the hard \gr\ emission}

The photon index of the high-energy component is extremely hard. In particular, it is harder than $\Gamma=1.5$, often adopted as a lower bound on the photon index, based on simple SSC models for the broad band emission from blazars. {Such a limit on the photon index of blazars underlies} the derivation of constraints on the EBL density (see e.g. \citet{1ES_nature,0229_HESS,1ESSS_HESS}). The detection of a hard spectral component in the flare of \mrk\ has, therefore, important implications for the modeling of blazar spectra and the EBL. In the following sections we discuss possible mechanism(s) for the production of a very hard spectral component.

\subsection{Mechanisms intrinsic to the source}
Several models which could explain very hard intrinsic blazar spectra in the \gr\ band have already been proposed.

\subsubsection{ Internal absorption origin}

 \citet{aharonian08} considered the possibility for the formation of a hard \gr\ blazar spectra through the absorption of \gr s in the dense soft photon background, with a narrow (e.g. thermal) spectrum. \gr s with energies $E_\gamma$ are most efficiently absorbed through interactions with soft photons of energy $\epsilon_s\simeq 10^2\left[ E_\gamma/10^{10}\mbox{ eV}\right]^{-1}$~eV. Optical depths $\tau\sim 1-10$ would be sufficient to produce a hard spectrum above the energy $10\left[\epsilon_s/10^2\mbox{ eV}\right]^{-1}$~GeV. The observation of a hard spectral component in \mrk\ would imply, therefore, the existence of a dense field of soft photons with energies $\epsilon>100$~eV in a compact \gr\ emission region. If the soft photon field has a thermal spectrum, its temperature should be $T> 10^6$~K, much higher than the typical temperature of accretion disks in Active Galactic Nuclei (AGN) accreting in a radiatively-efficient way. At the same time, this temperature range might not be unrealistic for radiatively-inefficient accretion flows, like those found in the Fanaroff-Riley I radio galaxies such as the nearby radio galaxy M~87 \citep{neronov07}. A  potential problem for the ``absorption feature'' explanation of the hard spectrum is that if the soft photon field has a narrow energy distribution, the suppression of the \gr\ signal should become small at energies much lower than $10\left[\epsilon_s/10^2\mbox{ eV}\right]^{-1}$~GeV. This means that the flaring component of the spectrum should reappear at the low-energy end of the LAT energy range, which is not the case. This difficulty could, however, be overcome if the soft photon distribution has a broad spectrum extending to energies much higher than $\epsilon\simeq 100$~eV. This might also be typical for the radiatively inefficient accretion flows where the emission in the 0.1-1~keV range is supposed to be produced via inverse Compton (IC) by the accretion flow electrons with a quasi-thermal distribution \citep{sibiryakov08}.

\subsubsection{ IC from hard spectrum electron origin}. 

\citet{katarzinski06} and \citet{bottcher08} proposed that a hard spectra could be produced via the IC mechanism if the spectrum of IC emitting electrons is hard enough. In particular, \citet{katarzinski06} consider the possibility of a low energy cut-off in the electron spectrum. In this case the spectrum of IC emission in the SSC model could have a photon index as low as $\Gamma=1/3$. \citet{bottcher08} consider the possibility of a hard spectrum for the case when the seed photons for IC scattering are provided by the Cosmic Microwave background (CMB) and the electron spectrum has a slope $dN_e/dE\sim E^{-\gamma}$ with $\gamma<2$. 
Each of these models carries with it a potential problem with relation to source variability.

On the one hand, in the model of \citet{katarzinski06}, an immediate effect of cooling is that an $E^{-2}$ type spectrum forms at energies below the low-energy cut-off, giving rise to a $\Gamma=1.5$ photon spectra. Furthermore, if the synchrotron cooling of the highest energy flux occurs in the Klein-Nishina regime, even softer, $\Gamma>1.5$, photon spectra are obtained. The cooling times of electrons emitting above 10~GeV are $t_{\rm cool}\approx 5~(0.3~{\rm G}/B)$~days. Thus only fields of $\sim 0.1$~G would allow the hard spectrum IC emission to live for long enough to explain the persistence of the hard spectrum during the 30~day flare. Furthermore, the acceleration time of 10~GeV electrons in such fields \citep{Rieger:2006md}, $t_{\rm acc.}\approx 10^{-5}~(0.1~c/v_{\rm sh})^{2}$~days, leaves the question as to the origin of the rise time of the flare unanswered. A possible way out of this difficulty is  that the flare rise time is determined not by the acceleration time scale, but by the dynamical time scale of the system (this would imply, however, that there is a pre-existing population of accelerated electrons). Cooling times shorter than the synchrotron cooling time $t_{\rm cool}$ could also be achieved if an additional cooling mechanism is present, such as adiabatic cooling \citep{finke08}.

Similarly, the problem of the cooling time for the model of \citet{bottcher08} limits its applicability to short flares. For this model, the cooling time of electrons emitting at $E_\gamma\sim 10-100$~GeV energies is very long, $t_{\rm IC}\simeq  4\times 10^5\left[E_\gamma/30\mbox{ GeV}\right]^{-1/2}$~yr, leaving this mechanism inapplicable for the description of a short flaring episode.

SSC type models embedded within the stochastic acceleration framework, however, do not suffer so badly from these timescale problems \citep{katarzinski06b}. The steady state spectra in this scenario, obtained when stochastic acceleration is balanced by radiative losses, can indeed be hard enough to explain the observed hard flare of Mrk~501. Furthermore, for these models, the longer acceleration time for 10~GeV electrons in 0.1~G magnetic fields, $t_{\rm acc.}\approx 9$~(100~km~s$^{-1}/\beta_{\rm A})^{2}$~days, can more naturally explain the rise-time of the flaring episode. The Alfven speeds motivated by this scenario, however, require rather dense astrophysical environments with $n\sim 10^{7}$~cm$^{-3}$ \citep{O'Sullivan:2009sc}.

\subsubsection{ Hadronic models}

Hard intrinsic spectra in the \gr\ range can also arise in hadronic models \citep{rachen98,aharonian00,mucke01,reimer04,zacharopoulou11}.  Variability times as short as $\sim 10^4$~s can originate from the synchrotron cooling time of protons in strong $\sim 100$~G magnetic fields, if the observed \gr\ emission has proton and/or muon synchrotron origin. A hard spectrum, within this framework, can form as the result of intrinsic absorption of the \gr\ flux on soft photon fields in the source \citep{zacharopoulou11}, in a way similar to the leptonic models with intrinsic absorption. Alternaviely, the spectrum of accelerated protons might be intrinsically hard and have a pileup at the highest energies \citep{aharonian00}.  Finally, if significant contribution to the source flux is produced through muon synchrotron emission, the hard spectrum could arise by muons with an energy below a certain critical energy decaying before they lose their energy through synchrotron emission \citep{rachen98,reimer04}. 

\subsubsection{ Vacuum gap acceleration origin}. 

The problem of the absence of cooling of electrons encountered in the models of \citet{katarzinski06} and \citet{bottcher08} disappears if one assumes that the observed \gr\ emission comes directly from the electron acceleration region. In this case the hard spectrum of electrons is maintained by the balance between the energy gain and energy loss rates. Such a possibility is considered e.g. in models describing particle acceleration in black hole magnetospheres \citep{beskin92,levinson00,neronov05, aharonian05,neronov07,neronov09} in which an almost monochromatic distribution of particles is maintained at the vacuum gap in the magnetosphere. The spectrum of synchrotron and/or curvature emission from electrons and/or protons accelerated in the vacuum gap can be very hard. For example, in the regime when the acceleration of particles of mass $m$ and charge $e$ in a magnetosphere with magnetic field $B$ is balanced by the synchrotron loss rate
$dE/dt\sim \kappa eB \sim -e^4B^2E^2/m^4=-P_{\rm synch}$
($\kappa\lesssim 1$ is acceleration efficiency), typical particle energies are 
$E\sim m^2\kappa^{1/2}/(e^{3/2}B^{1/2})$. The synchrotron radiation from such particles is sharply peaked at an energy $E_\gamma=eBE^2/m^3\sim \kappa m/e^2\simeq 10^2\kappa\left[m/m_p\right]\mbox{ GeV}$
independently of the magnetic field strength. The spectrum of emission below this peak energy is hard (see e.g. \citet{vincent10} for an example of the spectrum calculation). In principle, this hardness could be close to the limiting case for the synchrotron emission spectrum from a monochromatic electron distribution, which is characterized by a photon index $\Gamma=1/3$. If the energy loss mechanism limiting the maximal proton energies is curvature radiation, the maximal photon energy depends on the magnetic field and on the particle energy as $E_\gamma\simeq 30\left[(E/m)/10^{10}\right]^3\left[M/10^9\mbox{ cm}\right]^{-1}\mbox{ GeV}$,
where $M$ is the black hole mass which determines the size of the acceleration region. The \gr\ emission from the vacuum gap is highly anisotropic (see e.g. \citep{neronov07}). Most of the hard spectrum luminosity is directed along the AGN jet so that the \gr\ flux from the vacuum gap can be comparable or higher than the \gr\ emission from electrons accelerated in the jet. Since the energy deposited in the accelerated particles is released in the form of the \gr s \citep{neronov07}, the \gr\ production efficiency in the vacuum gap is extremely high.

\subsubsection{IC from electrons within a cold relativistic flow origin}. 

Alternatively, a narrow particle energy distribution could be maintained in a ``cold'' relativistic wind with very large bulk Lorenz factor, similar to the case of pulsar winds, if such winds are ejected by the central supermassive black hole. Hard spectrum \gr\ emission could then originate through IC emission, produced by the wind propagating through an external radiation field, as suggested by \citet{aharonian02}.
\vskip0.5cm

\subsection{Emission from \gr\ induced cascades in the intergalactic medium}
Very hard \gr\ emission spectra can also form through the propagation of multi-TeV \gr s from their source to Earth. The absorption of VHE \gr s through their interactions with the EBL initiates the development of electromagnetic cascades. If the energy of the primary \gr s is high enough, most of the primary source power is transferred into an electromagnetic cascade. The energy in the electromagnetic cascade is released in the form of detectable secondary cascade \gr\ emission with energies below the energy at which the mean free path of \gr s is comparable to the distance to the source \citep{coppi,neronov07a}. This results in the possibile formation of a hard \gr\ emission spectrum, resulting from the development of a cascade in the intergalactic medium, as was suggested by \citet{aharonian02}. An example calculation of the cascade spectrum in the case of \mrk\ is shown in Fig. \ref{fig:spectrum}. In this calculation primary \gr s with energy 100~TeV were injected at the source. The broad band $E^{-2}$ type spectrum in the energy range 0.1-10~GeV is formed during the propagation of the \gr\ beam towards the Earth. The observed high-energy cut-off at $E_{\rm cut}\simeq 5$~TeV sits at the energy for which the photon mean free path is about the distance to the source. 

The low energy suppression of the cascade spectrum arises due to the influence of intergalactic magnetic fields on the geometrical structure of the cascade. 
If the intervening intergalactic  magnetic field  (IGMF) strength were negligible, secondary $e^+e^-$ pairs, as well as cascade \gr s would propagate in the same direction as the primary \gr\ beam. Small deflections of the trajectories of electrons and positrons by weak, but non-negligible magnetic fields lead to the time delay of the cascade signal \citep{plaga95,japanese,neronov09a}. At energies for which the time delay is much longer than the flare duration, the cascade emission is suppressed.
This time delay increases with the decrease of the energy $E_\gamma$ of the cascade photons \citep{neronov09a}
$T_{\rm delay}=10^{1.5}\left[E_\gamma/10^2\mbox{ GeV}\right]^{-5/2}\left[B/10^{-17}\mbox{ G}\right]^{-2}
\mbox{ d.}$
The time delay suppression of the cascade flux becomes stronger at lower energies. This is illustrated in Fig.~\ref{fig:spectrum} where the cascade spectrum calculated for magnetic field values of $10^{-17}$~G, $3\times 10^{-17}$~G and $10^{-16}$~G are shown.  The low-energy suppression of the spectrum is observed below energies of $100$~GeV. Monte Carlo calculations which take into account the production spectrum of the pairs and of the secondary cascade photons show that at energies for which the time delay suppression is important, the spectrum of the cascade emission is hard, with a photon index $\Gamma\simeq 1$, close to the observed photon index of emission above 10~GeV in the Mrk 501 flare. Details of the Monte-Carlo code used for the calculation can be found in \citet{taylor11}.

The range of IGMF strengths required in the cascade model is consistent with the lower bounds on the IGMF which have been recently derived from the non-observation of the \gr\ emission from electromagnetic cascades emission from several hard spectrum TeV blazars \cite{neronov11,taylor11, tavecchio11,dermer11}. The exact  level of this lower bound depends on the nature of suppression of the cascade emission signal (time delay or extended nature of the cascade signal, see \citet{taylor11} for details) and is in the range of $10^{-17}-10^{-16}$~G if the IGMF correlation length is $\lambda_B\gg 1$~Mpc. For shorter correlation lengths, the bound strengthens to $B\gg 10^{-17}$~G. This cascade explanation of the observed hard spectrum of Mrk 501 is valid only if the real IGMF strength is close to the previously derived lower bounds.

The cascade model provides an alternative possible explanation of the extremely hard spectrum in the 10-200~GeV band. The interpretation of the data in terms of this model provides a measurement of the magnetic field in the intergalactic medium along the line of sight toward \mrk. The duration of the flare was $\sim 1$~month. If the magnetic field were weaker than $B\sim 10^{-17}$~G, the characteristic time delay in the energy range $E_\gamma\sim 100$~GeV would be much shorter than the flare duration and no suppression of the \gr\ flux would be observed (dotted curve in Fig. \ref{fig:spectrum}). On the other hand, if the magnetic field were stronger than $10^{-16}$~G, the $E^{-1}$ type spectrum formed by the time delay suppression effect would continue to energies much higher than $100$~GeV. This would place heavy demands on the source TeV luminosity, requiring a TeV flare luminosity of more than an order of magnitude higher than that observed in the 100~GeV energy band. Although this possibility can not be fully ruled out (no systematic monitoring of the source during the flare was done by the ground-based \gr\ telescopes), such a scenario would need a source flux larger than $3\times 10^{-10}$~erg/(cm$^2$~s), in excess of the highest historical flux detected from the source during the large 1997 flare (the spectrum of the 1997 flare measured by HEGRA telescope \citep{hegra} is shown in grey in Fig. \ref{fig:spectrum}). 

An alternative possibility for the cascade contribution to the Mrk~501 spectrum was considered by \citet{takahashi12}. Assuming that the observed TeV band flux is intrinsic to the source, rather than resulting from the development of an electromagnetic cascade, the cascade emission is still expected to appear in the GeV energy band. \citet{takahashi12} made an attempt to constrain the strength of IGMF along the line of sight towards Mrk 501 by searching for time delayed GeV band ``pair echo'' emission following the hard flare discussed in this paper. We note, however, that even if the peak flare flux in the TeV band was at the level measured by VERITAS, the Fermi data are of insufficient quality to provide sensible constraints on the IGMF. The reason for this is the high level of the steady-state flux from the source in the GeV energy band. From Fig. \ref{fig:lc} one can see that no variations in the flux level are detected in the 0.3-30~GeV band, neither during the flare, nor in the period following the flare. The source flux in the GeV band is constantly at the level of $EF_E\simeq 2\times 10^{-11}$~erg/cm$^2$s, which is above that expected for the pair echo flux at any moment of time (see Fig. 1 of  \citet{takahashi12}). Thus, the pair echo signal is not expected to be detectable, no matter how weak the magnetic field is. At the same time, if the peak flux of the flare was at the level indicated by the Fermi measurement in the $E>100$~GeV energy band (see Fig. \ref{fig:spectrum}), the strength of the pair echo signal considered by \citet{takahashi12} would be larger by a factor 3 -- 5, so that the cascade emission in the GeV band would be above the level of the steady state emission observed from the source.

\section{Conclusions}

The observation of a very hard spectrum of \gr\ emission from \mrk\ during an ``orphan'' \gr\ flare challenges theoretical models of \gr\ emission from blazars. Several mechanisms for the formation of the very hard spectra are able to provide explanations for the observed source behaviour.  To clarify the nature of the hard flare(s) a systematic analysis of the frequency of occurrence and properties of the hard and/or ``orphan'' \gr\ flares in the TeV \gr\ loud BL Lacs is needed.  
\vskip0.5cm

{\it Acknowledgements.} The work of AN and AMT is supported by the Swiss National Science Foundation grant PP00P2\_123426.

\end{document}